# FLEXIBLE VARIABLE STAR EXTRACTOR:
# new software for detection of variable stars

BREUS, VITALII [1]

(1) Department "Mathematics, Physics and Astronomy", Odessa National Maritime University, Mechnikova, 34, 65029 Odessa, Ukraine, bvv_2004@ua.fm

**Abstract:** We developed software for detection of variable stars using CCD photometry. It works with "varfind data" that could be exported after processing CCD frames using C-Munipack. Our goals were maximum automation and support of large fields of view with thousands of stars. The program chooses the comparison stars automatically, processes all time series using multiple comparison stars to get final light curves. Different filtering algorithms are used to reduce the impact of outlying points, imaging artifacts and low quality CCD frames without careful manual time series reduction. We implemented various variable detection indices and plotting two-channel diagrams of selected pair of indices and mean brightness of the star to distinguish variables from constant stars for further manual check of outlying points as variable candidates.

## Introduction

An indisputable advantage of the CCD photometry vs. earlier technologies is that CCD observations allow us to measure brightness of thousands stars from the same field of view simultaneously. We may learn from the experience, that there is at least one more known variable star within 10-20 arc minutes of any primary object of investigation and sometimes we may be lucky and discover an unknown one. Over the years of using CCD photometry different techniques for detect variable stars were developed. It has improved in recent years also due to intensified interest to discovery of extrasolar planets using transit photometry.

One of the simplest algorithms is based on the dependency of noise level to mean brightness of the star. According to the statistics, if all stars were constant, the dependence of standard deviation of brightness vs. mean brightness of an object would have a parabola-like shape. A variable star obviously should have larger standard deviation than a constant object of the same mean brightness. Particularly, this algorithm is implemented in the C-Munipack software package (Motl, 2011). It requires manual review of all suspected variable stars' light curves to approve or reject the candidate.

However, this solution is sensitive to outlying points and sometimes fails if the data is noisy due to various reasons. In this case at the dependence of standard deviation on the mean brightness we may see variable stars in a heap of points corresponding to constant stars, and many constant stars with lack of data or outlying points will be located above the curve like they are variables. One of the reasons of this software development was the necessity of the solution capable to analyze few experimental CCD series with more then 10000 stars in each field of view for the purpose of searching unknown short period variables.

This paper contains a brief description of Flexible Variable Star Extractor software (hereafter named FVSE) and methods used for detection of variable stars.

## Algorithms

We used the basic idea of classical analysis of standard deviation on the mean brightness dependence. Usually this chart is based on the differential magnitudes and user can make a decision if a star is variable or not. This way we discovered eclipsing binary 2MASS J18024395 + 4003309 (Andronov & Breus & Zola, 2012, Andronov et al., 2015). However, using of the artificial comparison star instead of the control star makes possible to increase accuracy estimates by a factor of 1.3-2.1 times for clear and cloudy nights, respectively (Kim & Andronov, 2004). The algorithm of the „mean weighted" comparison star for the photometry using CCD-cameras was described by Andronov & Baklanov (2004) and implemented in MCV. Obviously, using final magnitudes obtained by this method instead of differential ones for standard deviation calculation allows to decrease the noise and influence of the scatter of some particular comparison star on the data. It is computationally more expensive so it takes relatively more time to process the same data set, but it does not require modern PCs even for large number of stars and long time series.

During the process of developing it turned out that standard deviation does not distinguish variables significantly in case of big number of stars, so we decided to implement also other indices. Recently Sokolovsky et al. (2017) analyzed 18 statistical characteristics quantifying scatter and/or correlation between brightness measurements and compared their performance in identifying variable objects using time series obtained with different class telescopes. We choose Median absolute deviation, chi-squared, Robust median statistic, Normalized excess variance and von Neumann ratio based on performance and least computational expenses.

**Implementation**

FVSE reads magnitudes and error estimates from the "varfind" data exported from C-Munipack. The weighted mean magnitude and standard variance are calculated for every star in the data set where weight of each observation is inversely proportional to the square of the individual observation error estimates.

The selection of comparison stars may be done manually or automatically. FVSE chooses the comparison stars using experimentally derived criteria, which could be adjusted at the settings tab. First the star should not be close to the frame border because these data usually has lower quality (default value: 50 pixels). The comparison star should be measured at least at 70% of CCD frames of the time series. Its standard variance must not exceed the value of the mean variance of all stars multiplied by the factor of 1.5. This allows us to select less noisy stars as the comparisons. The total number of selected comparison stars is limited (default: 1000). Since the stars are ordered from bright to faint in varfind data, this cuts the faintest objects that usually do not fit one of previously mentioned criteria. For some data sets it's necessary to limit the number of comparison stars to traditional 10-15.

Following the algorithm of the „mean weighted" comparison star, weights of selected comparison stars are determined and the final magnitudes of all stars are calculated using the obtained artificial comparison star and the known magnitude of selected comparison star (entered by user). Since this moment, user could view light curves of any star and export it to the text file. Here the raw instrumental magnitudes is presented with red points, final light curve with green points (see Fig. 1).

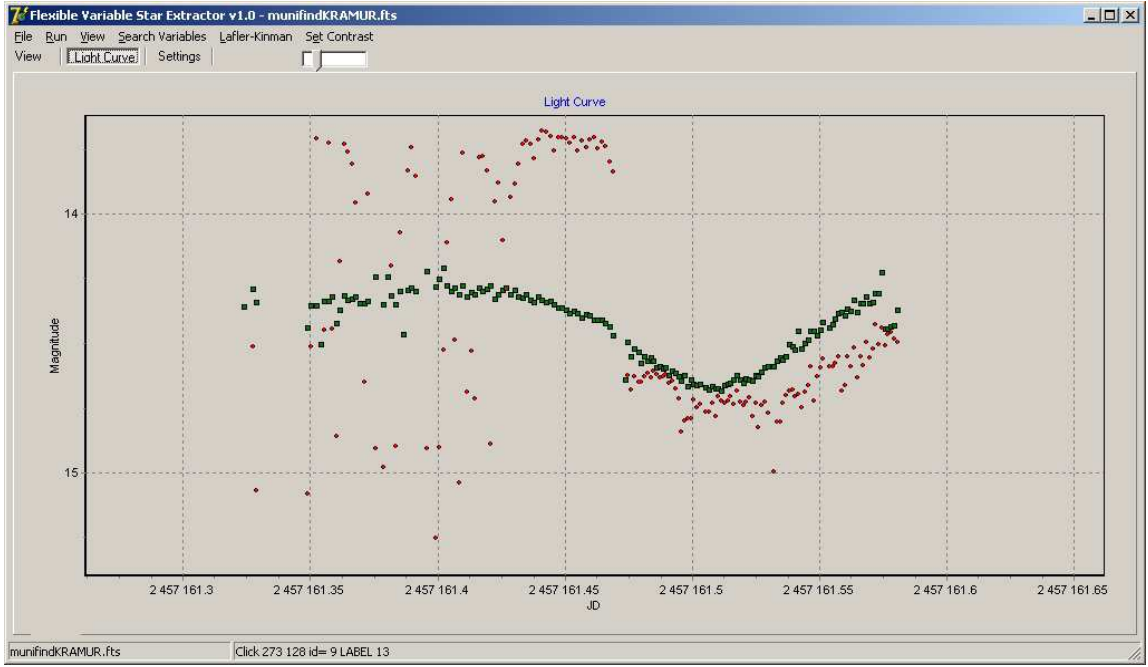

Figure 1: A sample light curve of the EW-type system V442 Cam during one night of observations (Breus et al., 2010). Large scatter at the beginning corresponds to the partially cloudy weather conditions that changed to clear sky later. The scatter of the final light curve (green) is also larger, but the influence of clouds is reduced using artificial comparison star method.

"Search Variables" menu item runs the calculation of variability detection indices used to distinguish variable stars from constant ones. We determine standard deviation $\sigma$, chi-squared $\chi^2$, Median absolute deviation, Robust median statistic, Normalized excess variance and von Neumann ratio. The chi-squared value is used in the form of square root and von Neumann ratio is calculated in inversed and simplified way. The detailed description of software and implementation of these algorithms was presented by Breus (2017).

User can plot any pair of standard deviation, mean brightness and these variability detection indices at the diagram. In any case, most of constant stars will form a group of points, and any outlying point may correspond to a variable star. The decision whether an object is a variable or not is left to the user who can view or save a light curve of any star in one click.

On this step it's necessary to mark the viewed star as variable or false alarm using context menu (see Fig. 2), this way it does not necessary to remember all objects in large field of view contrary to other solutions. At any step it is possible to remove some outlying points or even whole CCD frames from the data set and re-calculate the light curves and variability indices.

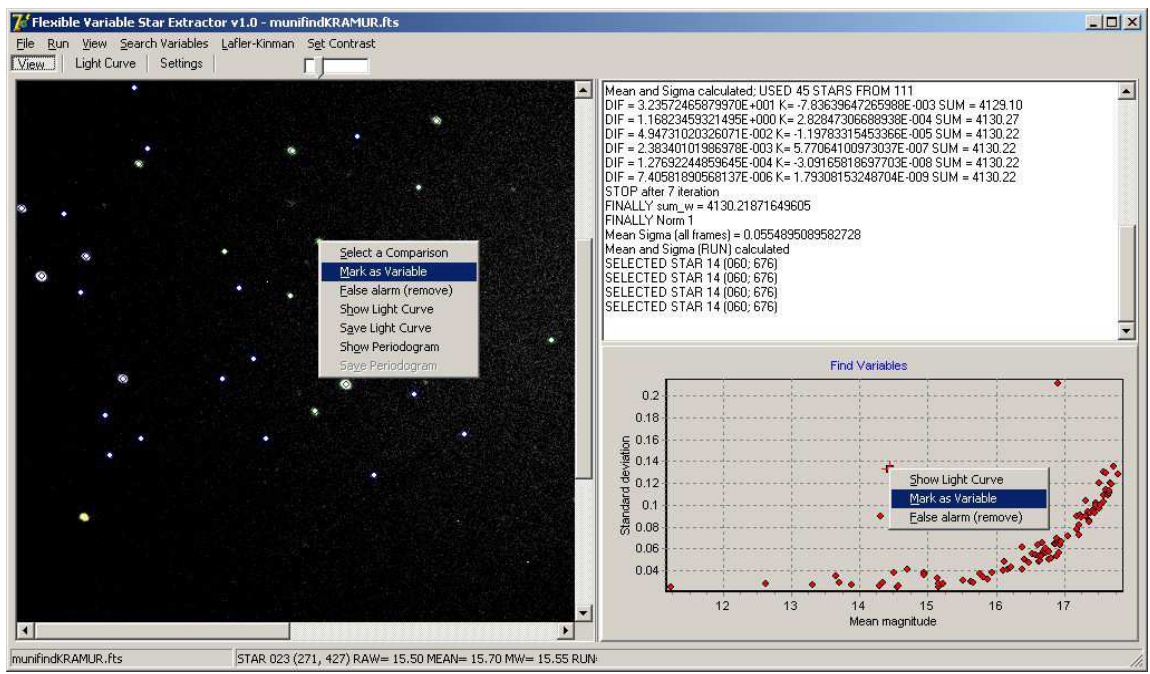

Figure 2: Graphical user interface of FVSE. Collage shows the context menu which may be called clicking on the star at the image or clicking the point at the diagram. It's possible to view the light curve of selected star, mark it as variable or false alarm.

It's possible to calculate periodograms of all stars in the data set. Originally there was an idea to use some parameters of the periodogram as one more variability detection index, but it has not yet been implemented. For this purpose, the Lafler – Kinman - Kholopov method was used (Lafler & Kinman, 1965, Andronov & Chinarova, 1997) and the implementation was borrowed from Variable Stars Calculator software (Breus, 2003, Breus, 2007).

Current version has hardcoded limits of the image size (5120x5120 pixels), 100000 points in the time series and 40000 stars in the field of view (may be changed upon request). This was done to increase performance in avoiding using dynamic arrays. These values are enough for most of the data and setting higher limits will raise RAM requirements.

FVSE is available as a freeware[1].

---

[1] http://uavso.org.ua/varsearch/

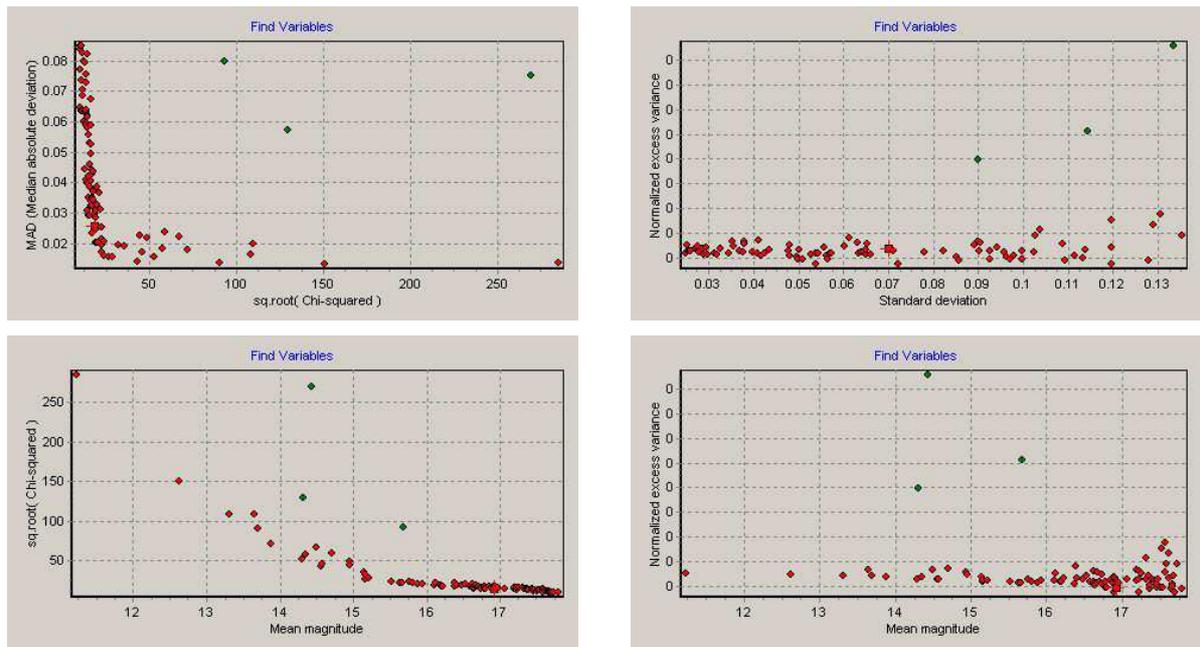

Figure 3: Samples of two-channel diagrams for the field of view of the intermediate polar MU Cam: dependence of MAD vs. chi-squared, NEV vs. standard deviation, chi-squared on the mean magnitude and NAV vs. mean magnitude. Green points correspond to MU Cam, V442 Cam and V440 Cam.

**Results and Conclusion**

We improved one of the most popular algorithms for variable stars detection in order to decrease the influence of noisy data and outlying points on the chart. "Flexible Variable Star Extractor" software was developed. It works with the data exported from C-Munipack, chooses the comparison stars automatically and processes all time series.

We implemented different variability detection indices. It turned out that some variability detection indices are effective to distinguish a star of particular variability type; other indices do not show significant difference for variable and constant stars, at the same time for another types of variability the situation is opposite. Innovative solution of viewing two-channel diagrams of any pair of parameters allows marking variable stars of all types by viewing 2 or 3 diagrams if the first one is not sufficient.

This style significantly improved the speed of semi-automatic search for variable stars in generic field of view of any size and time series of any duration.

Software was developed in the framework of the Ukrainian Virtual Observatory project (Vavilova et al., 2012). This program will be applied to the observations obtained within the "Inter-Longitude Astronomy" (Andronov et al., 2017) and "AstroInformatics" campaigns (Vavilova et al., 2017).